\documentclass[ twocolumn,aps,prd,   
               preprintnumbers,numbers,sort&compress,nofootinbib,
                            showpacs,
               colorlinks,
               linkcolor=blue,
               citecolor=blue]{revtex4-1}
   \newcommand{\exclude}[1]{}

\usepackage{graphicx,amsmath,amssymb,bm}
\usepackage{psfrag}
\usepackage{feynmp}
\usepackage{hyperref}
\usepackage{enumitem}
\newcommand{\nn}{\nonumber}
\newcommand{\beq}{\begin{equation}}
\newcommand{\eeq}{\end{equation}}
\newcommand{\be}{\begin{eqnarray}}
\newcommand{\ee}{\end{eqnarray}}
 \newcommand{\Lqcd}{\Lambda_{\mathrm{QCD}}}

\def\+{\dagger}

\def\ra{\rangle}
\def\<{\langle}
\def\>{\rangle}

\usepackage{epsfig}
\usepackage{dcolumn}

\begin{document}

\title{ Casimir scaling in gauge theories with a gap. Deformed QCD as a toy model.}

\author{Evan Thomas \& Ariel R. Zhitnitsky}
 \affiliation{Department of Physics \& Astronomy, University of British Columbia, Vancouver, B.C. V6T 1Z1, Canada}

\begin{abstract}

We study a Casimir-like behaviour in a ``deformed QCD''.
We demonstrate that for the system defined on a manifold of size $ \mathbb  L $ 
the difference  $\Delta E\equiv E -E_{\mathrm{Mink}}$  between the energies  of a system in a non-trivial background  and Minkowski space-time geometry exhibits the Casimir-like scaling 
 $\Delta E \sim \mathbb{L}^{-1}$,
despite the presence of a mass gap in the system, in contrast with naive expectation 
 $\Delta E \sim \exp(-m{\mathbb  L })$,
which would normally originate from any physical massive propagating degrees of freedom consequent to conventional dispersion relations. 
The Casimir-like behaviour in our system comes instead from a non-dispersive (``contact'') term which is not related to any physical propagating degrees of freedom, such that the naive argument is simply not applicable.
These ideas can be explicitly tested  in weakly coupled deformed QCD.   We comment on profound consequences for cosmology of this effect if it persists in real strongly coupled QCD. 
 
  \end{abstract}
\maketitle

\exclude{
We study a Casimir-like non-dispersive contribution to the topological susceptibility which exhibits the opposite sign as normal dispersive contributions due to propagating degrees of freedom. We discuss how this ``contact'' term is related not to locally acting physical degrees of freedom, but to the large scale boundary conditions imposed on the problem so that it is fundamentally related to the topological behaviour of the theory rather than the specific microscopic description. We state the known exact expressions for the topological susceptibility in a two dimensional $U(1)$ gauge theory, and derive similar expressions for a non-trivial ``deformed'' $SU(N)$ gauge theory in three dimensions, which despite exhibiting confinement, degeneracy of topological sectors, and non-trivial theta dependence, is weakly coupled and under full theoretical control. We argue that the presence of such a Casimir-like term in these theories suggests its presence in true QCD also, and gives precisely the type of dependence 
on the boundary conditions which is required for a dynamical model for the Dark Energy rooted in strongly coupled gauge theory.
}

\section{Introduction and motivation} \label{introduction}

The main motivation for the present studies is a recent suggestion on the dynamical Dark Energy (DE) model which is entirely rooted in the strongly coupled QCD, without any new fields and/or coupling constants ~\cite{dyn,4d,Zhitnitsky:2010ji,Zhitnitsky:2011tr}. The key element  of the proposal ~\cite{dyn,4d,Zhitnitsky:2010ji,Zhitnitsky:2011tr} is based on paradigm  that  the relevant energy which enters the Einstein equations is in fact the difference $\Delta E\equiv E -E_{\mathrm{Mink}}$ between the energies  of a system in a non-trivial background  and Minkowski space-time geometry,  similar to the well known Casimir effect when the observed energy is a difference between the energy computed for a system with conducting boundaries (positioned at finite distance $ \mathbb  L $) and infinite Minkowski space
\footnote{Here and in what follows we use term ``Casimir effect'' to emphasize the power like sensitivity to large distances irrespectively to  their nature. A crucial distinct feature which characterizes the system  we are interested in is the presence of dimensional parameter $ \mathbb  L \sim H^{-1}$ (where $H$ is a Hubble constant)  in the system which discriminates it  from infinitely large  Minkowski space-time.}. 
This paradigm is based on the conjecture that gravity, as described by the Einstein equations, is a low-energy effective interaction which, as such, should not be sensitive to the microscopic degrees of freedom in the system but to some effective scale. Thus, the energy density that enters the semiclassical Einstein equations should not be the ``bare'' energy as computed in QFT, but rather a ``renormalized'' energy density. We propose the renormalization scheme given above which sets the vacuum energy to zero in Minkowski space wherein the Einstein equations are automatically satisfied as the Ricci tensor identically vanishes.
 
The above prescription is in fact the standard subtraction procedure that is normally used for the description of horizon thermodynamics \cite{Hawking:1995fd,Belgiorno:1996yn} as well as in a course of computations of different Green's function in a curved background by subtracting infinities originated from the flat   space~\cite{Birrell:1982ix}. In the present context such a definition $\Delta E\equiv (E -E_{\mathrm{Mink}})$ for the vacuum energy was first advocated in 1967 by Zeldovich~\cite{Zeldovich:1967gd} who argued that  $\rho_{\text{vac}} \sim Gm_p^6 $ with $m_p$ being the proton's mass. Later on such a definition for the relevant energy $\Delta E\equiv (E -E_{\mathrm{Mink}})$ which  enters the Einstein equations has been advocated from   different perspectives in a number of papers, see, for example, the relatively recent works~\cite{Bjorken:2001pe, Schutzhold:2002pr, Klinkhamer:2007pe, Klinkhamer:2009nn, Thomas:2009uh, Polyakov:2009nq,Krotov:2010ma,Maggiore:2010wr} and references therein.

We study the scaling behavior of $\Delta E$ when the background deviates slightly from minkowski space. The difference $\Delta E$  must obviously vanish when any deviations (parametrized by Hubble constant or inverse size of the visible universe, $H\sim  \mathbb  L^{-1}$) go to zero as this corresponds to the transition to flat Minkowski space. A naive expectation based on common sense suggests that $\Delta E \sim \exp(-\Lqcd/H)\sim \exp(-10^{41})$ as QCD has a mass gap $\sim \Lqcd\sim 100~ {\text MeV}$, and therefore, $\Delta E$ must not be sensitive to size of our universe $ \mathbb  L \sim H^{-1}$. Such a naive expectation formally follows from the dispersion relations which dictate that a sensitivity to very large distances must be exponentially suppressed when  the mass gap is present in the system\footnote{The  Casimir effect  due to the massless $E\&M$ field  obvious shows such power dependence $\Delta E=-\frac{\pi^2}{720  \mathbb L^4}$.    Similar computations for a massive scalar particle with mass 
$m$   leads to an exponentially suppressed result $\Delta E\sim\exp (-m \mathbb  L )$ as expected, see e.g.\cite{Casimir}.}. 

However, as emphasized in \cite{Zhitnitsky:2010ji,Zhitnitsky:2011tr} in strongly coupled gauge theories along with conventional dispersive contribution there exists a non-dispersive contribution, not related to any physical propagating degrees of freedom.  This non-dispersive (contact) term generally  emerges as a result  of topologically nontrivial sectors in four dimensional QCD. The variation of this  contact term with variation of the background may lead to a power like scaling $\Delta E\sim H +{\cal O} (H)^2$  rather than to an exponential like $\Delta E \sim \exp(-\Lqcd/H)$.  
If true, the difference between two metrics (FLRW and Minkowski) would lead to an estimate 
   \be
   \label{Delta}
   \Delta E\sim \frac{\Lqcd^3}{ \mathbb  L }\sim (10^{-3} {\text eV})^4,~ 1/ \mathbb  L \sim H \sim 10^{-33} {\text eV}
   \ee
which is amazingly close to the observed DE value today.  It is interesting to note that 
expression (\ref{Delta})  reduces to Zeldovich's formula   $\rho_{\text{vac}} \sim Gm_p^6 $ if one replaces $ \Lambda_{QCD} \rightarrow m_p $   and $  H\rightarrow G \Lambda_{QCD}^3$. The last step follows from the solution of the Friedman equation 
\be \label{friedman}
  H^2=\frac{8\pi G}{3}\left(\rho_{DE}+\rho_M\right),  ~~ \rho_{DE}\sim H\Lqcd^3
\ee    
when the DE component dominates the matter component, $ \rho_{DE}\gg\rho_M$. In this case    the evolution of the universe  approaches a  de-Sitter state with constant expansion rate $H\sim G \Lambda_{QCD}^3$ as follows from (\ref{friedman}).
\exclude{
  A comprehensive phenomenological analysis of this model has been recently performed in ~\cite{Cai:2010uf}, see also \cite{Sheykhi:2011xz,RozasFernandez:2011je} where comparisons with current observational data including SnIa, BAO, CMB, BBN have been presented.  The conclusion was that this model is consistent with all presently available data. The main goal of this paper is not comparison of this model with observations; we refer the reader to ~\cite{Cai:2010uf} on this matter. Rather,  the main goal of this paper is to get some deep theoretical insights behind  the Casimir type behaviour (\ref{Delta}) in a gapped theory such as QCD.
}

Another motivation to study the Casimir like  behaviour in QCD is a proposal \cite{Zhitnitsky:2010zx, Zhitnitsky:2012im} that the $\cal{P}$ odd correlations observed at RHIC and LHC is in fact another manifestation of long range order advocated in this work. Furthermore, an apparently universal thermal spectrum observed in all high energy collisions when the statistical thermalization could never be reached in the systems, might be also related to the same contact term, not related to any physical propagating degrees of freedom, see \cite{Zhitnitsky:2010zx, Zhitnitsky:2012im} and references therein for the details.

There are a number of arguments supporting the power like behaviour $\Delta E\sim H +{\cal O} (H)^2$ in gauge theories, see section \ref{T} where we present some general arguments suggesting the Casimir like corrections in gauge theories with nontrivial topological structure. However, it is always desirable and very instructive to see how the general arguments work in some simplified settings. 
This is precisely the goal for the present study: we want to consider a simplified (``deformed'') version of QCD which, on one hand, would be a weakly coupled gauge theory wherein computations can be performed in theoretically controllable manner. On other hand, this deformation would preserve all the relevant elements of strongly coupled QCD such as confinement, degeneracy of topological sectors, nontrivial $\theta$ dependence, presence of non-dispersive contribution to topological susceptibility, and other crucial aspects, for this phenomenon to emerge. Such a ``deformed'' theory has recently been developed \cite{Yaffe:2008}. All computations in this work (excluding those in sec. \ref{instantons}) are performed within this framework.  
  
\section{Topological susceptibility in the deformed QCD}\label{section-chi}
In the deformed theory an extra term is put into the Lagrangian in order to prevent the center symmetry breaking that characterizes the QCD phase transition between ``confined" hadronic matter and ``deconfined" quark-gluon plasma. Thus we have a theory which remains confined at high temperature in a weak coupling regime, and for which it is claimed \cite{Yaffe:2008} that there does not exist an order parameter to differentiate the low temperature   confined regime from the high temperature confined regime. The non-trivial topological sectors of the theory are described in this model in terms of the weakly coupled monopoles. The monopoles in this framework are not real particles, they are pseudo-particles which live in Euclidean space and describe the physical tunnelling processes between different topological sectors $|n\ra$ and $| n+1 \ra$. In particular, the monopole fugacity $\zeta$    should be understood as number of tunnelling events per unit time per unit volume
\be
\label{zeta}
  \left(\frac{  {\rm number ~of  ~tunnelling ~ events}}{VL}\right)=\frac{N_c\zeta}{L},
\ee
where extra factor $N_c$ in (\ref{zeta})  accounts for $N_c$ different types of monopoles present in the system and  $L$ is the size of the circle along $\tau=i t$ and plays the role of the inverse temperature.  The monopole gas experiences Debye screening so that the field due to any static charge falls off exponentially with characteristic length $m_{\sigma}^{-1}$. The number density $\cal{N}$ of monopoles is given by the monopole fugacity, $\sim \zeta$, so that the average number of monopoles in a ``Debye volume" is given by
\begin{equation} \label{debye}
{\cal{N}}\equiv	m_{\sigma}^{-3} \zeta = \left( \frac{g}{2\pi} \right)^{3} \frac{1}{\sqrt{L^3 \zeta}} \gg 1,
\end{equation} 
The last inequality holds since the monopole fugacity is exponentially suppressed, $\zeta \sim e^{-1/g^2}$, and in fact we can view (\ref{debye}) as a constraint on the validity of the  approximation.

The topological susceptibility in this model can be explicitly computed and is given by \cite{Thomas:2011ee}
\be \label{YM}
  \chi_{YM} = \int d^4 x \< q(\bold{x}) q(\bold{0}) \>	=\frac{\zeta}{N_cL} \int d^3 x \left[ \delta(\bold{x}) \right].
\ee
The light quarks can be easily inserted into the system. The corresponding generalization of eq. (\ref{YM}) reads \cite{Thomas:2011ee}
\be \label{chi_QCD}
  \chi_{QCD} &=& \int d^4 x \< q(\bold{x}) q(\bold{0}) \>  \\
    &=&\frac{\zeta}{N_cL} \int d^3 x \left[ \delta(\bold{x})-m_{\eta'}^2 \frac{e^{-m_{\eta'}r}}{4\pi r}  \right].  \nn
\ee
The first term in this equation has  non-dispersive nature, similar to eq. (\ref{YM})  and   has the positive sign. This contact term (which is not related to any physical propagating degrees of freedom) has been computed  in this model using  monopoles describing the transitions between the degenerate topological sectors
\footnote{
  In the context of this paper the ``degeneracy'' implies there existence of winding states $| n\ra$ constructed as follows: ${\cal T} | n\ra= | n+1\ra$.  In this formula the operator ${\cal T}$ is  the  large gauge transformation operator  which commutes  with the Hamiltonian $[{\cal T}, H]=0$. The physical vacuum state is {\it unique} and constructed as a superposition of $| n\ra$ states. In QFT approach the presence of $n$ different sectors in the system is reflected  by  summation over $ { n \in \mathbb{Z}}$ in  definition of the path integral in QCD. It should not be confused with conventional term ``degeneracy'' when two or more physically {\it distinct} states are present in the system.
}. 
The positive sign of this term is the crucial element of the resolution of the $U(1)_A$ problem. The second term in eq. (\ref{chi_QCD}) is a standard dispersive  contribution, can be restored through the absorptive part using conventional dispersion relations, and has a negative sign  in accordance with general principles. This conventional physical contribution is saturated in this model by the lightest $\eta'$ field. It enters $\chi_{QCD}$ precisely in such a way that the Ward Identity (WI) expressed as $\chi_{QCD} (m_q=0)=0$ is automatically satisfied as a result of cancellation between the two terms. If the contact non-dispersive term with ``wrong sign'' was not present in the system, the WI could not be satisfied as physical states always contribute with negative sign in eq. (\ref{chi_QCD}). 

One should note that the number of tunnelling events per unit time per unit volume (\ref{zeta}) in pure gauge  theory in this model (with no  quarks) precisely  concurs  with the absolute value of the energy density of the system. Furthermore,   the topological susceptibility in pure gauge  theory calculated  as the second derivative  of $E_{\mathrm{YM}}(\theta) $ with respect to $\theta$   precisely coincides with non-dispersive contact term with ``wrong sign'' explicitly and directly computed in (\ref{chi_QCD}). Indeed,  
\be \label{vac}
  E_{\mathrm{YM}}(\theta)&=&- \frac{N_c\zeta}{L}\cos\left(\frac{\theta}{N_c}\right), \\ 
  \chi_{YM} (\theta=0)&=&  \left. \frac{\partial^2E_{\mathrm{YM}}(\theta)}{\partial \theta^2}
    \right|_{\theta=0}=\frac{\zeta} {N_cL}, \nn
\ee
where we keep only the lowest branch $l=0$ in expression for  $\cos\left(\frac{\theta +2\pi l}{N_c}\right)$ to simplify formula (\ref{vac}), see detail discussions with complete set of references on this matter in \cite{Thomas:2011ee}. In other words, the contact term in pure gauge theory $ \chi_{YM}  =\frac{\zeta}{N_cL}$ can be interpreted in terms of number tunnelling events between different topological sectors in the system. Therefore, there is no surprise that it has the ``wrong sign" as the relevant physics can not be described in terms of propagating physical degrees of freedom, but rather, is described in terms of the tunnelling events between  different (but physically equivalent) topological sectors in the system.

\section{Casimir-type  behaviour   in   deformed QCD  }\label{Casimir}
Up to this point the theory  was formulated on  ${\mathbb R}^3\times S^1$
with small compactification size $L$ for compact time coordinate $S^1$ and infinitely large  space  ${\mathbb R}^3$
describing  three other dimensions. As explained in section \ref{introduction}, we are actually  interested in behaviour of the system when a space with large dimensions ${\mathbb R}^3$ receives some small modifications,
for example the theory is defined in a   ball  ${\mathbb R}^3\rightarrow {\mathbb B}^3$ with ${\mathbb L}$ being a very large  size of the compact dimension of the sphere ${\mathbb S}^2$ which is a boundary of the ball ${\mathbb B}^3$. Such a modification can be thought as a simplest way to model and test the sensitivity of our theory to arbitrary large distances such as size of our visible universe determined by the Hubble constant $H/\Lqcd\sim 10^{-41}$. We want to know how the topological susceptibility of the system which describes the $\theta$ dependent portion of the vacuum energy $E_{\mathrm{vac}}(\theta=0)$ changes with slight variation of size of the system. We assume that ${\mathbb L}\sim H^{-1}\sim 10 ~{\text Gpc}$ is  much larger than any other scales of the problem. Essentially we want to see whether our deformed QCD model with a mass gap $m_{\sigma}$   predicts an  exponential scaling  typical  for a free massive particle 
\be \label{naive}
  \Delta E ({\mathbb L})\equiv\left[E ({\mathbb B}^3)-E({\mathbb R}^3)\right]   \sim \exp(-m_{\sigma}{\mathbb L})  
\ee
or, it demonstrates  a Casimir type behaviour 
\be \label{casimir}
  \Delta E ({\mathbb L})\equiv \left[E ({\mathbb B}^3)-E({\mathbb R}^3)\right] \sim \frac{1}{\mathbb L}
     + {\cal O}\left(\frac{1}{\mathbb L}\right)^2.
\ee
If we did not have a non-dispersive contribution in our system, we would   immediately predict the behaviour  (\ref{naive}) as the only available option for a gapped theory in close analogy  with  conventional Casimir computations    for a massive particle $\Delta E ({\mathbb L})\sim \exp(-m  {\mathbb L})$, see e.g. review paper~\cite{Casimir}.  However, our system is much richer, more complicated, and more interesting, as it exhibits a non-dispersive term resulting from degeneracy of topological sectors in gauge theory as discussed in the text. This contact term, being unrelated to any physical degrees of freedom, may provide different scaling properties since conventional dispersion relations do not dictate its behaviour at very large distances. As we shall argue below, the deformed QCD indeed exhibits the Casimir type behaviour (\ref{casimir}) in drastic contrast with the conventional viewpoint represented by eq.(\ref{naive}). As we reviewed in section \ref{introduction} we interpret a tiny deviation of 
the $\theta$- dependent vacuum energy $E_{\mathrm{vac}} $ in expanding universe (in comparison with Minkowski space-time) as a main source of the observed dark energy.  The Casimir type behaviour (\ref{casimir})  plays a key role in possibility of such an identification. 

We start our discussions in section \ref {instantons}  with conventional 4d instanton computations \cite{'tHooft:1976fv} in which infrared regularization for some gauge modes is required and achieved by putting the system into a sphere with finite radius $\mathbb L$.  It allows us to compute  power like corrections to the standard instanton density \cite{'tHooft:1976fv}. However, the corresponding  corrections  being computed for a fixed instanton size $\rho$  can not be interpreted  as a physically observable quantity  because the integral $\int d\rho$   over large size instantons diverges for this system
when semiclassical approximation for large $\rho$ breaks down. Nevertheless, this example explicitly shows when and why a Casimir type correction 
(to conventional formula computed in infinite  ${\mathbb R}^4 $ space)  emerges. 

Next, we compute a similar correction for the ``deformed QCD'' model in section \ref{monopoles} wherein a Casimir type correction also appears, resulting from the same physics related to topological sectors of the theory. In contrast with the previous case, the correction computed in this system is physically ``observable'' quantity as it represents the vacuum energy of the system. Indeed, the tunnelling transitions in this case are described by weakly coupled monopoles, such that semiclassical computations  of the vacuum energy (\ref{zeta}),(\ref{YM}) expressed in terms  of the density $\zeta$ of pseudo-particles are fully justified.  The size of pseudo-particles (fractionally charged monopoles) which describe the tunnelling events in this model is fixed by construction see \cite{Yaffe:2008} \cite{Thomas:2011ee} for the details.

\subsection{Casimir-type corrections for 4d instantons }\label{instantons}

Our goal here is to study a power like correction to the instanton density described in the classic paper \cite{'tHooft:1976fv}.
As such, we adopt 't Hooft's notation, and in particular, use the same background-dependent gauge $C_4={\cal D_{\mu}}A_{\mu}^{\text{a qu}}$, which drastically simplifies all computations.
Essentially, the problem is reduced to analysis of the normalization factors for finite number of zero modes (8 for $SU(2)$ gauge group) in this gauge wherein the system is defined in a sphere with large but finite radius radius $\mathbb L$.
Essentially we follow the construction described in section XI of ref. \cite{'tHooft:1976fv}. The corresponding  normalization factor explicitly enters the expression for the instanton density as it accompanies the integration over collective variables.
The contribution from non-zero modes does not exhibit such corrections, see the few comments on this issue at the end of this section.
We now concentrate on the zero modes and power like corrections which accompany the normalization factors if the system is defined on a large but finite space ${\mathbb B}^4_{\mathbb{L}}$ (four dimensional interior of a ball of radius ${\mathbb L}$) rather than an infinite space ${\mathbb R}^4$.

We start with four translational zero modes which have the form
\be \label{translations}
  A_{\mu}^{\text{a qu}} (\nu)\sim \eta_{a\mu\nu}(1+r^2)^{-2},~~ \nu =1, ..., 4
\ee
where we literally use  't Hooft's notations for $ \eta_{a\mu\nu}$ symbols and dimensionless coordinate $r^2=x_{\mu}^2$ measured in units of $\rho=1$. Computing the corresponding correction factor due to the translation zero modes $\kappa_{\rm tr.}$, we have
\be \label{kappa_1}
  \kappa_{\rm tr.}\equiv\frac{ \int^{\mathbb L}_0 d^4x [A_{\mu}^{\text{a qu}} (\nu)]^2}{\int^{\infty}_0 d^4x [A_{\mu}^{\text{a qu}} (\nu)]^2}\simeq
      \left[1-\frac{3}{\mathbb L^4}+{\cal O} (\frac{1}{\mathbb L^6}) \right].    
 \ee
The corresponding correction factor to the instanton density has power like correction as anticipated. As a result of additional rotational symmetry one should expect, in general,  $ \mathbb L^{-2}$ corrections, while translation zero modes lead to a much smaller correction $\sim \mathbb L^{-4}$ as eq. (\ref{kappa_1}) shows. It will be neglected in what follows. Dilaton and global gauge rotations lead to 
$\sim \mathbb L^{-2}$ as we discuss below. 

For the dilaton zero mode 
\be \label{dilaton}
  A_{\mu}^{\text{a qu}} \sim \eta_{a\mu\nu}x^{\nu}(1+r^2)^{-2} 
\ee
a similar formula reads
\be \label{kappa_2}
  \kappa_{\rm dil.}\equiv\frac{ \int^{\mathbb L}_0 d^4x [A_{\mu}^{\text{a qu}} (\nu)]^2}{ \int^{\infty}_0 d^4x [A_{\mu}^{\text{a qu}} (\nu)]^2}\simeq
      \left[1-\frac{3}{\mathbb L^2}+{\cal O} (\frac{1}{\mathbb L^4}) \right], 
 \ee
such that the correction to the instanton density is proportional to $\sqrt{\kappa_{\rm dil.}}\simeq (1-\frac{3}{2\mathbb L^2})$.
  
Computing the corresponding contribution due to three  zero modes related to global gauge rotations requires much more refined analysis as explained in  \cite{'tHooft:1976fv}. This is due to the specific features of the background dependent gauge $C_4={\cal D_{\mu}}A_{\mu}^{\text{a qu}}$ when the corresponding three modes are pure gauge artifact. As shown  in  \cite{'tHooft:1976fv}  the corresponding contribution is finite, but very sensitive to the infrared regularization determined by the size \exclude{$ \mathbb L$} $R$ of large sphere. The corresponding contribution to the instanton density is $\sim (\lambda_4 V)^{3/2} $ where $V$ is the four volume, while $\lambda_4\sim V^{-1}$ is defined as follows
\be
 \label{lambda}
  \lambda_{4}&=&\frac{\int_Vd^4x[\psi^a_{\mu}(b)]^2}{\int_Vd^4x[\psi^a(b)]^2}, ~~ b=1,2,3, \\
   \psi^a(b) &=& \eta_{a\mu\nu}\bar{\eta}_{b\mu\lambda}\frac{x^{\nu}x^{\lambda}}{(1+x^2)}, \nonumber \\
   \psi^a_{\mu}(b)&=&{\cal D_{\mu}} \psi^a(b) = \eta_{a\lambda\mu}\bar{\eta}_{b\lambda\nu}\frac{x^\nu}{(1+x^2)^2}. \nonumber 
\ee
The corresponding power like corrections can be computed in a similar manner to the other zero modes, except that we must retain the regularization since the denominator above diverges as $\sim V$. So we have the two correction factors
\be
  \kappa_{num.} \equiv \frac{\int_0^{\mathbb L} d^4x[\psi^a_{\mu}(b)]^2}{\int_0^\infty d^4x[\psi^a_{\mu}(b)]^2}
   \simeq \left[1-\frac{3}{\mathbb L^2}+{\cal O} (\frac{1}{\mathbb L^4}) \right], \nonumber
\ee
and
\be
  \kappa_{den.} \equiv \frac{V(R)}{V({\mathbb L})}\frac{\int_0^{\mathbb L} d^4x[\psi^a(b)]^2}{\int_0^R d^4x[\psi^a(b)]^2}
   \simeq \left[1-\frac{4}{\mathbb L^2}+{\cal O} (\frac{1}{\mathbb L^4}) \right]. \nonumber
\ee
The fraction, $V(R)/V({\mathbb L})$, is the correction to $V$ in the instanton density factor, and is included here so that we can take the regularization $R\rightarrow \infty$. The combined gauge rotation correction factor is then   
 \be
 \label{kappa_3}
 \kappa_{\rm rot.} \equiv \frac{\kappa_{num.}}{\kappa_{den.}} \simeq
  \left[1+\frac{1}{\mathbb L^2}+{\cal O} (\frac{1}{\mathbb L^4}) \right],
 \ee
such that the correction to the instanton density is proportional to $ (\kappa_{\rm rot.})^{3/2}\simeq (1+\frac{3}{2\mathbb L^2})$. Accidentally,  for $SU(2)$ gauge group  the leading $\mathbb L^{-2}$ correction from the  dilation (\ref{kappa_2})  and global gauge rotations (\ref{kappa_3})  exactly cancel each other. This accidental cancellation does not hold for general $SU(N)$ gauge group  when power of $\kappa_{\rm rot.}$ enters the instanton density with a different power. 

We remark here that the technique used in \cite{'tHooft:1976fv} is essentially a variational approach wherein the boundary conditions are implemented implicitly rather than explicitly. It allows us to use all the zero modes (\ref{translations}),(\ref{dilaton}),(\ref{lambda}) as well as standard classical instanton solution in the original form defined on $\mathbb R^4$ in which the conformal invariance is a symmetry of the system.
So in this approach, neither the instanton itself, nor its zero modes (\ref{translations}),(\ref{dilaton}),(\ref{lambda}) are solutions of the equation of motions which vanish at the boundary. 
This approach has been tested in many follow up papers, and we adopt it in the present work using the same technique in the next section.
We also point out that the conformal invariance is explicitly broken in the one instanton sector by the size of the instanton $\rho$, such that corrections take the form $ (\frac{\rho^2}{\mathbb L^2})^n$.
It is restored by the integration $\int d\rho$. 
However, in this paper we are interested in by the computation in one instanton sector only when dimensional parameter $\rho$ is explicitly present in the system, and it is small and fixed. 

The important message here is that such kind of power correction do appear in general.
The source of these corrections is a long range tail of zero modes.
We can not derive a definite conclusion from these computations because the integral over large size instantons $\int d\rho$ diverges and the semiclassical approximation breaks down.  
However, the same problem studied in the deformed QCD model considered in Section \ref{monopoles} does not suffer from such deficiencies as semiclassical computations are under complete theoretical control.
Thus, a Casimir like correction to the monopole fugacity $\zeta$ in this model is explicitly translated to the correction to the vacuum energy density and topological susceptibility (\ref{vac}), supporting (\ref{casimir}) and in huge contrast with naive expectation (\ref{naive}).
It is important that the source of the corrections in the deformed QCD model is the same as in undeformed QCD considered here, and that source is the long range tails of the zero modes, which lead to large distance sensitivity.
The only difference is that the role of the instanton size $\rho$ in computations above in the one instanton sector is played by the inverse monopole's mass $m_W^{-1}$ in next section \ref{monopoles}.
Because it is a true scale of the problem however, $m_W^{-1}$ is not integrated over as $\rho$ is.
   
\subsection{Casimir-type corrections for 3d monopoles }\label{monopoles}

We now turn to the deformed gauge theory described in \cite{Yaffe:2008, Thomas:2011ee} wherein the low-energy behaviour is given by a $U(1)^N$ Coulomb gas of monopoles in Euclidean ${\mathbb R}^3$. Basically, we want to understand the dependence of the monopole fugacity, $\zeta$, which comes out of the measure transformation to collective coordinates, on the size of the system, ${\mathbb L}$. In this case, as in the previous section, we consider the interior of a sphere of large but finite radius ${\mathbb L}$. There are four zero-modes present in this system: three translations since the monopoles are in ${\mathbb R}^3$, no dilations since the monopole size is fixed by the symmetry-breaking scale in this model $m_W$, and one gauge rotation since the gauge group for a given monopole is $U(1)$.  As in \cite{'tHooft:1976fv}, we work in a regular gauge to remain sensitive to the large distance physics. The monopole solution in the ``hedgehog'' regular gauge is given by
\be \label{hedgehog}
  v_\mu^a(x) &=& \epsilon_{\mu\nu a}\frac{x^\nu}{\left| x \right|^2}\left[ 1-\frac{m_W \left| x \right|}{\sinh\left(m_W \left| x \right|\right)}\right], \\
  \phi^a(x) &=& \frac{x^a}{\left| x \right|^2}\left[ m_W\left| x \right|\coth\left(m_W\left| x \right|\right)-1\right],
\ee
 where we adapted notations from \cite{Dorey:1997ij,Davies:2003} treating the monopole measure in supersymmetric Yang -Mills theory. 
In formula (\ref{hedgehog})     $v_\mu^a$ denotes the three spacial gauge fields for the classical solution, and $\phi^a$ the gauge field in the compact time direction (the ``Higgs'' field in this model) when all fields can be combined in a single 4d field $v_m$.   

We then want to   compute the correction factors for the collective coordinate measure coming from these four zero modes when the system is defined in a large but finite sphere. We closely follow the 't Hooft's  treatment  \cite{'tHooft:1976fv} presented in previous section~\ref{instantons}. We start by considering the translation modes defined by the spacial derivative of the classical monopole solution (\ref{hedgehog})
\be \label{translation_mon}
  Z_m^a(\nu) =  - {\partial_\nu}v_m^a(x-z)  + {\cal D}_{m} v_{\nu}^a =v_{m\nu}^a
\ee
where second term on the right hand side of eq. (\ref{translation_mon}) is necessary to keep $ Z_m^a(\nu)$ in the background gauge, see 
\cite{Dorey:1997ij,Davies:2003} for the details. 
This leads us to the following expression for  correction factor due to the translation zero modes
\be \label{kappa_4}
  \kappa_{\rm tr.} \equiv\frac{ \int^{\mathbb L}_0 d^4x [ Z_m^a(\nu)]^2}{ \int^{\infty}_0 d^4x [ Z_m^a(\nu)]^2}
   \simeq \left[1-\frac{1}{m_W {\mathbb L}} + {\cal O} (\frac{1}{\mathbb L^2})\right]
\ee 

Next we consider the gauge rotation zero-mode. As in the previous section, the contribution to the collective coordinate measure, and so the monopole fugacity, is $\sim(\lambda V)^{\frac{1}{2}}$ where $V$ is the three-volume and $\lambda$ is given by
\be \label{lambda_mon}
  \lambda &=& \frac{\int_V d^3 x\left[B^a_{\mu}\right]^2}{\int_V d^3 x[\phi^a]^2} \\ 
   B^a_\mu &=& \frac{1}{2}\epsilon_{\mu\nu\rho}\partial_\nu v^a_\rho = {\cal D_{\mu}}\phi^a. \nonumber
\ee
Again, the denominator diverges as $\sim V$ and we look at the two correction factors
\be \label{kappa_num}
  \kappa_{\rm num.} \equiv \frac{\int_0^{\mathbb L} d^3x[B^a_{\mu}]^2}{\int_0^\infty d^3x[B^a_{\mu}]^2} 
   \simeq \left[1-\frac{1}{m_W {\mathbb L}} + {\cal O} (\frac{1}{\mathbb L^2})\right], \nonumber
\ee
and
\be \label{kappa_den}
  \kappa_{\rm den.} \equiv \frac{V(R)}{V({\mathbb L})}\frac{\int_0^{\mathbb L} d^3x[\phi^a]^2}{\int_0^R d^3x[\phi^a]^2} 
   \simeq \left[1-\frac{3}{m_W {\mathbb L}} + {\cal O} (\frac{1}{\mathbb L^2})\right]. \nonumber
\ee
The total correction factor for the gauge rotation mode is then
\be \label{kappa_5}
  \kappa_{\rm rot.}\equiv\frac{\kappa_{\rm num.}}{\kappa_{\rm den.}}
   \simeq \left[1+\frac{2}{m_W {\mathbb L}} + {\cal O} (\frac{1}{\mathbb L^2})\right],
\ee
and therefore the total correction to the monopole fugacity from the (\ref{lambda_mon}) is $\sqrt{\kappa_{\rm rot.}}\simeq(1+\frac{1}{\mathbb L})$. Assembling the total correction to the fugacity,
\be \label{fugacity_correction}
   {\kappa}^{3/2}_{\rm tr.} {\kappa^{1/2}_{\rm rot.}} \simeq \left[1-\frac{1}{2 m_W {\mathbb L}} + {\cal O} (\frac{1}{\mathbb L^2})\right].
\ee

Thus, the deformed QCD, when put on a manifold with a boundary, receives some corrections to the monopole fugacity compared to Minkowski space that are power-like in the manifold size. The correction (\ref{fugacity_correction}) to the monopole fugacity leads immediately to the same correction to the topological susceptibility and so the background energy density through the relation (\ref{vac}). To be more precise, 
\be \label{final}
  \zeta(\mathbb L)=\zeta  \cdot\left[1-\frac{1}{2 m_W {\mathbb L}} + {\cal O} (\frac{1}{\mathbb L^2})\right],
\ee
where $\zeta$ is the monopole fugacity which enters the relation (\ref{vac}) computed in infinite Minkowski space.
We emphasize that the energy density changes in the bulk of space-time, not only in the vicinity of the boundaries, similar to the Casimir effect when the bulk energy density changes as a result of merely presence of the boundary. To reiterate, the deformed QCD, despite the presence of a mass gap, displays a suprising Casimir-like sensitivity to large distance boundaries, such that the energy density differs from the Minkowski space value by $\Delta E\sim \frac{1}{m_W {\mathbb L}}$. Again, this is in contrast to the naive expectation based on analysing the physical degrees of freedom, $\Delta E\sim e^{-m {\mathbb L}}$ with $m\sim m_{\sigma}$ being the lowest mass scale of the problem (\ref{naive}). 

\subsection{A few general comments}\label{comments}

Computations of the Casimir corrections presented above were based on an analysis of the zero modes when corresponding normalization factor explicitly enters the instanton/monopole density.
Now, we want to present some arguments suggesting that a corrections due to the non-zero mode contributions can be neglected, and, therefore, it cannot cancel the zero modes contribution.
Indeed, the computation of non-zero mode contribution is reduced to analysis of the phase shifts in the scattering matrix which can not change the normalization of the wave function itself as the only changes occur are the phase shifts.
An absolute normalization is dropped from the final formula for the instanton/monopole density when the ratio of the eigenvalues is considered.
This argument is consistent with observation that non-zero mode contribution depends on matter context of the theory as it varies when massive scalar of spinor fields in different representations are part of the consideration.
At the same time, the Casimir type corrections computed above are exclusively due to the gauge portion of the theory, not its matter context.
Indeed, these Casimir corrections were derived in pure gluodynamics.
So, it is difficult to imagine how a Casimir correction to non-zero mode contribution (if it is nonzero) may cancel a Casimir type correction originated from analysis of gauge zero modes.

We also comment that the correction $ \mathbb L^{-1} $ occurs as a manifestation of a slow power like decay of the zero modes in the background of a topologically nontrivial gauge configuration.
It should be contrasted with conventional behaviour of zero modes with a mass gap present in the system from the very beginning (for example, the well studied problem of a double well potential).
In former case, the zero modes decay according to the power law and leads to the Casimir type correction, while in the later case, the zero modes are well localized configurations which decay exponentially fast at large distances and can not be sensitive to large distance physics.
The mass gap is present for all physical degrees of freedom in both models.
However, in the former case the mass gap emerges as a result of the same instanton/monopole dynamics, while in the later a mass gap was present in the system from the very beginning and it was not associated with any instanton/monopole dynamics.
QCD obviously belongs to the former case, and we therefore expect this effect will persist in real strongly coupled QCD.

Next, our computations of the Casimir correction to the instanton/monopole density are based on assumption of the dilute gas approximation.
This is enforced in section \ref{instantons} by a finite instanton size $\rho$ which is kept fixed and small. 
On other hand, the semiclassical approximation in section \ref{monopoles} is automatically justified due to parametrically small fugacity $\zeta$, and total neutrality in this system is automatically achieved as long as the size of the system $ \mathbb L$ is much larger than the Debye screening length $m_{\sigma}^{-1}$, see (\ref{debye}).
In other words, we assume $\mathbb L \gg m_{\sigma}^{-1}$ such that neutrality of the system is automatically satisfied with exponential accuracy.
The finite size of the manifold does not spoil this neutrality if condition $ \mathbb L \gg m_{\sigma}^{-1}$ is satisfied.
Furthermore, the computation of the monopole's fugacity $\zeta$ and corresponding corrections (\ref{final}) can be performed without taking into account of the interaction of a monopole with other particles from the system as it would correspond to higher order corrections in density expansion $\sim\zeta^2$.
This is precisely the procedure which was followed in the original computations by Polyakov in \cite{Polyakov:1976fu} and in the deformed QCD model in ref.\cite{Yaffe:2008} at weak coupling.

Also, we emphasize that in variational approach developed in \cite{'tHooft:1976fv} neither the classical solution nor the corresponding zero modes vanish at the boundary of a finite size manifold.
The constraints related to the finite size $\mathbb L$ of the manifolds are accounted for implicitly rather than explicitly in this approach.
In particular, one should not explicitly cut off the classical action of the configuration as a result of finite size $\mathbb L$ where instanton/monopole is defined as this contribution is implicitly taken into account by variational approach.
However, even if we use an explicit cutoff for classical solution (as some people suggest) it still cannot cancel the zero mode corrections as these terms have different behaviour in $N$.
Indeed, the correction to the classical solution would be one and the same for any $N$, while corrections due to zero modes depend on $N$ as correction (\ref{kappa_5}) counts number of gauge rotations for $SU(N)$ gauge theory. 

Finally, it is quite possible that we overlooked some other possible corrections (for example, some corrections due to the boundaries which may occur in close vicinity of these boundaries).
We emphasize that our main result is not a computation of a specific coefficient in front of the correction to fugacity in equation (\ref{final}).
Rather, our main point is that these type of corrections do occur in a system with a gap, and it is very difficult to imagine that some boundary corrections may mysteriously cancel these computed bulk corrections (as some people suggest). 
Therefore, we present below some arguments and examples suggesting that Casimir type behaviour in gauge theories is in fact quite generic, rather than a peculiar feature of the system.

\section{Topological sectors  and  the Casimir correction in QCD}\label{T}

In this section we want to present few generic arguments suggesting that the emergence of a Casimir-like behaviour is not an accident, and not a computational error.
Rather, the effect has a deep theoretical roots as argued in \cite{Zhitnitsky:2011aa}.
We review these arguments starting with analogy with the well known Aharonov-Casher effect as formulated in \cite{Reznik:1989tm}.
The relevant part of that work can be stated as follows.
If one inserts an external charge into superconductor when the electric field is exponentially suppressed $\sim \exp(-r/\lambda)$ with $\lambda $ being the penetration depth, a neutral magnetic fluxon will be still sensitive to an inserted external charge at arbitrary large distance.
The effect is purely topological and non-local in nature.
The crucial point is that this phenomenon occurs, in spite of the fact that the system is gapped, due to the presence of different topological states in the system.
We do not have a luxury of solving a similar problem in strongly coupled four dimensional QCD analytically.
However, one can argue that the role of the ``modular operator''  of \cite{Reznik:1989tm} (which is the key element in the demonstration of long range order) is played by large gauge transformation operator $\cal{T}$ in QCD which also commutes with the Hamiltonian $[{\cal T},H]=0$, such that our system must be transparent to topologically nontrivial pure gauge configurations, similar to transparency of the superconductor to the ``modular electric field", see \cite{Zhitnitsky:2011aa} for the details. 
 
We interpret the computational results in a number of systems where Casimir like corrections have been established as a manifestation of the same physics which can be described in terms of the operator $\cal{T}$. We should mention that there are a few other systems, such as topological insulators, where a topological long range order emerges in spite of the presence of a gap in the system. We refer to ref \cite{Zhitnitsky:2011aa} for relevant references and details.  
  
There are a number of simple systems in which the Casimir type behaviour $\Delta E\sim \mathbb L^{-1} +{\cal O} (\mathbb L)^{-2}$ has been explicitly computed.
In all known cases this behaviour emerges from non-dispersive contributions when the dispersion relations do not dictate the scaling properties of this term.
 
The first example is an explicit computation~\cite{Urban:2009wb} in exactly solvable two-dimensional QED defined in a box size $\mathbb L$.
The model has all elements crucial for present work: non-dispersive contact term which emerges due to the topological sectors of the theory.
This model is known to be a theory of a single physical massive field.
Still, one can explicitly compute  $\Delta E \sim \mathbb L^{-1} $ in contrast with naively expected exponential suppression, $\Delta E\sim e^{-\mathbb L}$~\cite{Urban:2009wb}.
Another piece of support for a power like behaviour is an explicit computation in a simple case of Rindler space-time in four dimensional QCD ~\cite{Zhitnitsky:2010ji, Zhitnitsky:2010zx, ohta} where Casimir like correction have been computed using the unphysical Veneziano ghost which effectively describes the dynamics of the topological sectors and the contact term when the background is slightly modified.
Thus, power-like behaviour is not a specific feature of two dimensional physics as some people (incorrectly) interpret the results of ~\cite{Urban:2009wb}.
 
Our next example is  2d $CP^{N-1}$ model formulated on finite interval with size $\mathbb L$ \cite{Milekhin:2012ca}. 
In this case one can explicitly see emergence of $\Delta E \sim \mathbb L^{-1}$  in large $N$ limit in close analogy to our case (\ref{final}) where a theory has a gap, but nevertheless, exhibits the power like corrections.
The correction computed in \cite{Milekhin:2012ca} also comes from a non-dispersive contribution which can not be associated with any physical propagating degrees of freedom, similar to our case (\ref{final}). 

Power like behaviour $\Delta E \sim \mathbb L^{-1}$ is also supported by recent lattice results \cite{Holdom:2010ak}. The approach advocated in ref.\cite{Holdom:2010ak} is based on physical Coulomb gauge, in which nontrivial topological structure of the gauge fields is represented by the so-called Gribov copies leading to a strong infrared singularity.
Thus, the same Casimir-like scaling emerges in a different framework where the unphysical Veneziano ghost (used in refs. \cite{Zhitnitsky:2010ji, Zhitnitsky:2010zx, ohta}) is not even mentioned. 

The very same conclusion also follows from the holographic description of the contact term presented in \cite{Zhitnitsky:2011aa}.
The key element for this conclusion follows from the fact that the contact term in holographic description is determined by massless Ramond-Ramond (RR) gauge field defined in the bulk of 5-dimensional space.
Therefore, it is quite natural to expect that massless R-R field in holographic description leads to power like corrections when the background is slightly modified. 

To avoid any confusion with terminology we follow \cite{Zhitnitsky:2011aa} and call this effect as ``Topological Casimir Effect'' where no massless degrees of freedom are present in the system, but nevertheless, the system itself is sensitive to arbitrary large distances.
It is very different from conventional Casimir effect where physical massless physical photons are responsible for power like behaviour.
From the holographic viewpoint discussed in \cite{Zhitnitsky:2011aa} the ``Topological Casimir Effect'' in our physical space-time can be thought as conventional Casimir effect in multidimensional space when massless propagating R-R field in the bulk is responsible for this type of behaviour, although this field is not a physical asymptotic state in our four dimensional world. 

\section{Conclusion and future directions}\label{conclusion} 

We tested a sensitivity of a deformed QCD model with non-trivial topological features to arbitrary large distances. 
A naive expectation based on dispersion relations dictates that a sensitivity to very large distances must be exponentially suppressed (\ref{naive}) when the mass gap is present in the system.
However, we argued that along with conventional dispersive contribution there exists a non-dispersive contribution, not related to any physical propagating degrees of freedom. This non-dispersive (contact) term  with the ``wrong sign''  emerges as a result of topologically nontrivial sectors, and can be explicitly computed in our model.
The variation of this contact term with variation of the background leads to a power like ``Topological Casimir Effect'' (\ref{casimir}) in accordance with other arguments presented in section \ref{T} and in contrast with the naively expected exponential suppression (\ref{naive}). 

The ``Topological Casimir Effect"  in QCD, if confirmed by future analytical and numerical studies, may have profound consequences for understanding of the expanding FLRW universe we live in. 
We already mentioned in section \ref{introduction} that the observed DE (\ref{Delta}) may is fact be just a manifestation of this ``Topological Casimir Effect"  without adjusting any parameters.
In the adiabatic approximation the universe expansion can be modeled as a slow process in which the size of the system adiabatically  depends on time $\mathbb L(t)$ which leads to extra energy as equations (\ref{casimir}, \ref{final}) suggest. 
Such a model is obviously consistent with observations if $\mathbb L(t)$ is sufficiently large \cite{Urban:2009ke}.
We do not insist that this is the model of our universe.
Rather, we claim that if the effect persists in strongly coupled QCD, the energy density which can not be identified with any physical propagating degrees of freedom, is sensitive to arbitrary large distances as a result of nontrivial topological features of QCD.
Different geometries (such as FLRW universe) obviously would lead to different coefficients.
Nonetheless, the important message from these computations in our simplified model is that the energy density in the bulk is sensitive to arbitrary large distances comparable with the visible size of the universe, and that this sensitivity comes not from any new physics but simply from the proper treatment of the topological structure of QCD. 

We add that a comprehensive phenomenological analysis based on this idea has been recently performed in ~\cite{Cai:2010uf} where comparison with current observational data including SnIa, BAO, CMB, BBN has been presented, see also \cite{ohta,Sheykhi:2011xz,RozasFernandez:2011je,Cai:2012fq, Saaidi:2012av,Feng:2012wx} with related discussions.
The conclusion was that the model (\ref{Delta}) is consistent with all presently available data, and we refer the reader to these papers on analysis of the observational data.

Finally, what is perhaps more remarkable is the fact that the ``Topological Casimir Effect'' which is the subject of this work can be, in principle, experimentally tested in heavy ion collisions, where a similar environment can be achieved, see  \cite{Zhitnitsky:2010zx,Zhitnitsky:2012im} for the details.
In particular, the $\cal{P}$ odd correlations observed at RHIC and LHC have been interpreted in  \cite{Zhitnitsky:2010zx,Zhitnitsky:2012im} as a result of long range order represented by the ``Topological Casimir Effect''.
  
\section*{Acknowledgements}

ARZ is thankful to  Dima Kharzeev, Larry McLerran, and other members of Nuclear Physics groups at BNL and  Stony Brook U.     for useful and stimulating discussions related to the subject of the present work.
This research was supported in part by the Natural Sciences and Engineering Research Council of Canada.
         


\begin{thebibliography}{99}

\bibitem{dyn}
  F.~R.~Urban and A.~R.~Zhitnitsky,
  Nucl.\ Phys.\  B {\bf 835}, 135 (2010)
  [arXiv:0909.2684 [astro-ph.CO]].
  \bibitem{4d}
  F.~R.~Urban and A.~R.~Zhitnitsky,
  Phys.\ Lett.\  B {\bf 688}, 9 (2010)
  [arXiv:0906.2162 [gr-qc]].
\bibitem{Zhitnitsky:2010ji}
  A.~R.~Zhitnitsky,
  Phys.\ Rev.\  {\bf D82}, 103520 (2010).
  [arXiv:1004.2040 [gr-qc]].

\bibitem{Zhitnitsky:2011tr}
  A.~R.~Zhitnitsky,
  Phys.\ Rev.\  {\bf D84}, 124008   (2011).
  [arXiv:1105.6088 [hep-th]].


\bibitem{Hawking:1995fd}
  S.~W.~Hawking and G.~T.~Horowitz,
  Class.\ Quant.\ Grav.\  {\bf 13}, 1487 (1996)
  [arXiv:gr-qc/9501014].

\bibitem{Belgiorno:1996yn}
  F.~Belgiorno and S.~Liberati,
  Gen.\ Rel.\ Grav.\  {\bf 29}, 1181 (1997)
  [arXiv:gr-qc/9612024].
 
\bibitem{Birrell:1982ix}
  N.~D.~Birrell and P.~C.~W.~Davies, {\it Quantum Fields In Curved Space}, Cambridge Univ.\ Pr.\ , 1982.

\bibitem{Zeldovich:1967gd}
  Y.~B.~Zeldovich,
  JETP Lett.\  {\bf 6}, 316 (1967)
  [Pisma Zh.\ Eksp.\ Teor.\ Fiz.\  {\bf 6}, 883 (1967)].


\bibitem{Bjorken:2001pe}
  J.~Bjorken,
  arXiv:hep-th/0111196.

\bibitem{Schutzhold:2002pr}
  R.~Schutzhold,
  Phys.\ Rev.\ Lett.\  {\bf 89}, 081302 (2002).

\bibitem{Klinkhamer:2007pe}
  F.~R.~Klinkhamer and G.~E.~Volovik,
  Phys.\ Rev.\  D {\bf 77}, 085015 (2008)
  [arXiv:0711.3170 [gr-qc]].

\bibitem{Klinkhamer:2009nn}
  F.~R.~Klinkhamer, G.~E.~Volovik,
  Phys.\ Rev.\  {\bf D79}, 063527 (2009).
  [arXiv:0811.4347 [gr-qc]].
  
\bibitem{Thomas:2009uh}
  E.~C.~Thomas, F.~R.~Urban, A.~R.~Zhitnitsky,
  JHEP {\bf 0908}, 043 (2009).
  [arXiv:0904.3779 [gr-qc]].

\bibitem{Polyakov:2009nq}
  A.~M.~Polyakov,
  Nucl.\ Phys.\  {\bf B834}, 316-329 (2010).
  [arXiv:0912.5503 [hep-th]].
  
\bibitem{Krotov:2010ma}
  D.~Krotov, A.~M.~Polyakov,
[arXiv:1012.2107 [hep-th]].
 
\bibitem{Maggiore:2010wr}
  M.~Maggiore,
  Phys.\ Rev.\  {\bf D83}, 063514 (2011).
  [arXiv:1004.1782 [astro-ph.CO]].

  
  \bibitem{Casimir}
  G.~Plunien, B.~Muller and W.~Greiner
  Phys.\ Rep.\  {\bf 134}, 87 (1986).
  
 
\bibitem{Zhitnitsky:2010zx}
  A.~R.~Zhitnitsky,
  Nucl.\ Phys.\  {\bf A853}, 135-163 (2011).
  [arXiv:1008.3598 [nucl-th]].
 
\bibitem{Zhitnitsky:2012im} 
  A.~R.~Zhitnitsky,
  Nucl.\ Phys.\ A {\bf 886}, 17 (2012)
  [arXiv:1201.2665 [hep-ph]].

  
  \bibitem{Yaffe:2008}
  M.~\"{U}nsal and L.~G.~Yaffe,
  Phys.\ Rev.\ D {\bf 78}, 065035 (2008).
  [arXiv:0803.0344 [hep-th]].

\bibitem{Thomas:2011ee} 
  E.~Thomas and A.~R.~Zhitnitsky,
  Phys.\ Rev.\ D {\bf 85}, 044039 (2012)
  [arXiv:1109.2608 [hep-th]].
  
\bibitem{'tHooft:1976fv} 
  G.~'t Hooft,
  Phys.\ Rev.\ D {\bf 14}, 3432 (1976)
  [Erratum-ibid.\ D {\bf 18}, 2199 (1978)].
  
\bibitem{Dorey:1997ij} 
  N.~Dorey, V.~V.~Khoze, M.~P.~Mattis, D.~Tong and S.~Vandoren,
  Nucl.\ Phys.\ B {\bf 502}, 59 (1997)
  [hep-th/9703228].
  
\bibitem{Davies:2003}
  N.~M.~Davies, T.~J.~Hollowood and V.~V.~Khoze
  J.\ Math.\ Phys.\ {\bf 44}, 3640 (2003).
  [arXiv:hep-th/0006011] 
 
\bibitem{Polyakov:1976fu} 
  A.~M.~Polyakov,
  Nucl.\ Phys.\ B {\bf 120}, 429 (1977).
  
\bibitem{Zhitnitsky:2011aa} 
  A.~R.~Zhitnitsky,
  Phys.\ Rev.\ D {\bf 86}, 045026 (2012)
  [arXiv:1112.3365 [hep-ph]].
  
\bibitem{Reznik:1989tm}
  B.~Reznik, Y.~Aharonov,
  Phys.\ Rev.\  {\bf D40}, 4178-4183 (1989).
 
 
\bibitem{Urban:2009wb}
  F.~R.~Urban, A.~R.~Zhitnitsky,
  Phys.\ Rev.\  {\bf D80}, 063001 (2009).
  [arXiv:0906.2165 [hep-th]].

\bibitem{ohta}  
  N.~Ohta,
  Phys.\ Lett.\  B {\bf 695}, 41 (2011)
  [arXiv:1010.1339 [astro-ph.CO]].
 
\bibitem{Milekhin:2012ca} 
  A.~Milekhin,
  arXiv:1207.0417 [hep-th].
  
\bibitem{Holdom:2010ak}
  B.~Holdom,
  Phys.\ Lett.\  {\bf B697}, 351-356 (2011).
  [arXiv:1012.0551 [hep-ph]].

\bibitem{Urban:2009ke} 
  F.~R.~Urban and A.~R.~Zhitnitsky,
  JCAP {\bf 0909}, 018 (2009)
  [arXiv:0906.3546 [astro-ph.CO]].
   
\bibitem{Cai:2010uf} 
  R.~-G.~Cai, Z.~-L.~Tuo, H.~-B.~Zhang and Q.~Su,
  Phys.\ Rev.\ D {\bf 84}, 123501 (2011)
  
   
  
\bibitem{Sheykhi:2011xz} 
  A.~Sheykhi and M.~Sadegh Movahed,
  Gen.\ Rel.\ Grav.\  {\bf 44}, 449 (2012)
  [arXiv:1104.4713 [hep-th]].
  
\bibitem{RozasFernandez:2011je} 
  A.~Rozas-Fernandez,
  Phys.\ Lett.\ B {\bf 709}, 313 (2012)
  [arXiv:1106.0056 [astro-ph.CO]].
   
\bibitem{Cai:2012fq} 
  R.~-G.~Cai, Z.~-L.~Tuo, Y.~-B.~Wu and Y.~-Y.~Zhao,
  arXiv:1201.2494 [astro-ph.CO].

\bibitem{Saaidi:2012av} 
  K.~Saaidi, A.~Aghamohammadi and B.~Sabet,
  arXiv:1203.4518 [physics.gen-ph].
  
\bibitem{Feng:2012wx} 
  C.~-J.~Feng, X.~-Z.~Li and P.~Xi,
  arXiv:1204.4055 [astro-ph.CO].
  


 


  
 
  
  
 
 
\end{thebibliography}
\end{document}